\documentclass[10pt,a4paper,twocolumn]{article}
\usepackage[english]{babel}
\usepackage[utf8]{inputenc} 
\usepackage[T1]{fontenc}
\usepackage{fancyhdr}
\usepackage{textcomp}
\usepackage[dvips]{epsfig}
\usepackage{amscd}
\usepackage{amssymb}
\usepackage{amsthm}
\usepackage{latexsym}
\usepackage{mathrsfs}
\usepackage{amsmath}
\usepackage{amsfonts}
\usepackage{amssymb}
\usepackage{graphicx}
\usepackage{caption}
\usepackage{siunitx}
\usepackage{subcaption}
\usepackage{enumitem}
\usepackage{color}
\usepackage[toc,page]{appendix}
\usepackage{authblk}
\usepackage{tikz}
\usepackage{color}
\usepackage{soul}
\soulregister\cite7
\soulregister\ref7
\usepackage{geometry}
\usepackage{physics}
\usepackage{fancyhdr}
\fancypagestyle{plain}{%
  \fancyhf{} 
  \fancyfoot[C]{\tiny ©2022. This manuscript version is made available under the CC-BY-NC-ND 4.0 license https://creativecommons.org/licenses/by-nc-nd/4.0/}
  
}

\usetikzlibrary{positioning,decorations,arrows,patterns,shapes,backgrounds}

\theoremstyle{plain}
\newtheorem{thm}{Theorem}
\theoremstyle{plain}

\theoremstyle{definition}
\newtheorem{defi}[thm]{Definition}
\newtheorem{rem}{Remark}
\providecommand{\keywords}[1]{\textbf{\textit{Keywords---}} #1}

\newcommand{\ie}{\emph{i.e.}~}

\DeclareMathOperator{\sgn}{sgn}

\DeclareMathOperator{\rang}{rank}

\usepackage[T1]{fontenc}
\usepackage{textcomp}
\RequirePackage[bookmarks,%
                colorlinks,%
                urlcolor=blue,%
                citecolor=blue,%
                linkcolor=blue,%
                hyperfigures,%
                pagebackref,%
                pdfcreator=LaTeX,%
                breaklinks=true,%
                pdfpagelayout=SinglePage,%
                bookmarksopen=true,%
                bookmarksopenlevel=2]{hyperref}

\newcommand{\R}{\ensuremath{\mathbb{R}}}

\title{Nonlinear Three-Tank System Fault Detection and Isolation \\ Using Differential Flatness}

\author[b]{Rim Rammal}
\author[,a]{Tudor-Bogdan Airimitoaie\thanks{Corresponding author: \texttt{tudor-bogdan.airimitoaie@ims-bordeaux.fr}}}

\author[a]{Pierre Melchior}
\author[a]{Franck Cazaurang} 

\affil[a]{Univ. Bordeaux, Bordeaux INP, CNRS, IMS, 33405 Talence, France}
\affil[b]{LAAS - CNRS, Université de Toulouse, CNRS, INSA,
UPS, 31400 Toulouse, France}

\date{}

\definecolor{Tudor}{RGB}{173,9,169}
\definecolor{Rim}{RGB}{173,9,169}

\begin{document}

\twocolumn[
  \begin{@twocolumnfalse}
    \maketitle
  \end{@twocolumnfalse}
 
]
 
  \begin{abstract}
Fault detection and isolation on hydraulic systems are very important to ensure safety and avoid disasters. In this paper, a fault detection and isolation method, based on the flatness property of nonlinear systems, is experimentally applied on the three-tank system, which is considered as a popular prototype of hydraulic systems. Specifically, fault indicators, called residues, are generated using flat output measurements, and for the purpose of fault isolation, a definition of the isolability is introduced. This definition allows the characterization of flat outputs that are useful for fault isolation. A sensitivity analysis is proposed in order to improve the robustness of the method. Multiplicative faults are considered on sensors and actuators.
    \end{abstract} 
    

\section{Introduction}
\label{section1}
The three-tank system is considered as a representative process of the aircraft fuel tank system, used by researchers to test various fault detection and isolation (FDI) methods. The fuel storage in the tanks and fuel consumption have a significant impact on the CG position of the aircraft \cite{langton2010aircraft}. In particular, the position of the center of gravity (CG) of an aircraft is very important for its stability and safety. Therefore, controlling fuel levels in each tank is essential to providing the desired CG position and guarantee the safety of the aircraft. To do this, it is important to have correct information about the fuel level in each tank and to have fuel flow supervision. Accordingly, any fault on the aircraft fuel system sensors or actuators may affect the aircraft's CG control and cause disasters. For this purpose, the application of a FDI technique on the aircraft fuel tank system is important to detect and isolate faults on sensors and actuators. 

The three-tank system has been considered previously for validation of FDI methodologies. The authors in \cite{khan2010robust} tested a robust fault detection filter in order to detect and isolate faults on the three-tank sensors and actuators, and in \cite{shields2000assessment}, a nonlinear observer has been designed, based on the nonlinear model of the three-tank system, in order to detect a leakage from a pipe. 

FDI methods are classically based on the notion of redundant measurements which can be obtained either by multiple sensors or analytical components generating fault indicators, called residues. They represent the gap between each measurement (physical or analytical). For survey papers on FDI see \cite{zhou2014overview,thirumarimurugan2016comparison}. In the ideal case of noise free observations, if all the residues are equal to zero, then there is no fault on the system. However, if at least one residue is different from zero, then a fault is detected. In practice, due to the presence of noise on the system, the residues are compared next to fixed thresholds. If at least one residue exceeds its threshold, then a fault is detected, otherwise, there is no fault on the system. Studies on tuning thresholds can be found in \cite{ding2008model,khan2009threshold}.

Recently, FDI methods based on the flatness property of nonlinear systems have also been shown to be effective in detecting and isolating faults on sensors and actuators (see  \cite{mai2007flatness,suryawan2010fault,martinez2014flatness}).
Roughly speaking, we recall that a nonlinear system is said to be flat if there exists a variable $z$, called flat output, such that all the system states, inputs and outputs can be expressed in function of $z$ and a finite number of its successive time derivatives. In \cite{mai2007flatness}, the flatness-based FDI method is used to estimate actuator faults only. The developed method in \cite{suryawan2010fault} is applied to linear systems and takes into account only sensor faults. In \cite{martinez2014flatness}, the proposed flatness-based FDI method can be applied on both linear and nonlinear systems and takes into account sensor and actuator faults. In this method, the measurement of the flat output is used to calculate the redundant variables. The fault detection is common with the other FDI methods: if a residue exceeds its threshold, then a fault is detected. However, the isolability is more complex and depends on the chosen flat output, and sometimes multiple flat outputs are needed to isolate all faults \cite{torres2013faultdetection}. Nevertheless, the choice of these flat outputs is not arbitrary, \ie there are flat outputs that, when used together, increase the isolability of faults and others that do not.

Recently, in \cite{rammal2020choice}, an open-loop characterization of the flat outputs for FDI has been briefly presented, without being evaluated. It allows the definition of flat outputs that are independent, which is useful for the isolability. The contribution of this paper is twofold:
\begin{itemize}
    \item[(i)] Firstly, the flatness based FDI method presented in \cite{rammal2020choice} is improved by including a sensitivity analysis of the residues with respect to the faults in order to improve its robustness. 
    \item[(ii)] Secondly, an experimental evaluation on a three-tank benchmark in the presence of PI feedback controller with a flatness based feedforward action is provided.
\end{itemize}

The paper is organized as follows: Section~\ref{section2} describes the benchmark. The flatness-based fault detection and isolation is presented in Section~\ref{section3}. Section~\ref{section4} presents the extension of the characterization of the flat outputs in closed-loop system. The experimental results are given in Section~\ref{section5}. Finally, Section~\ref{section6} concludes the paper.

\section{Description of the Three-Tank System}
\label{section2}

\begin{figure}[ht]
\centering
\includegraphics[width=\columnwidth]{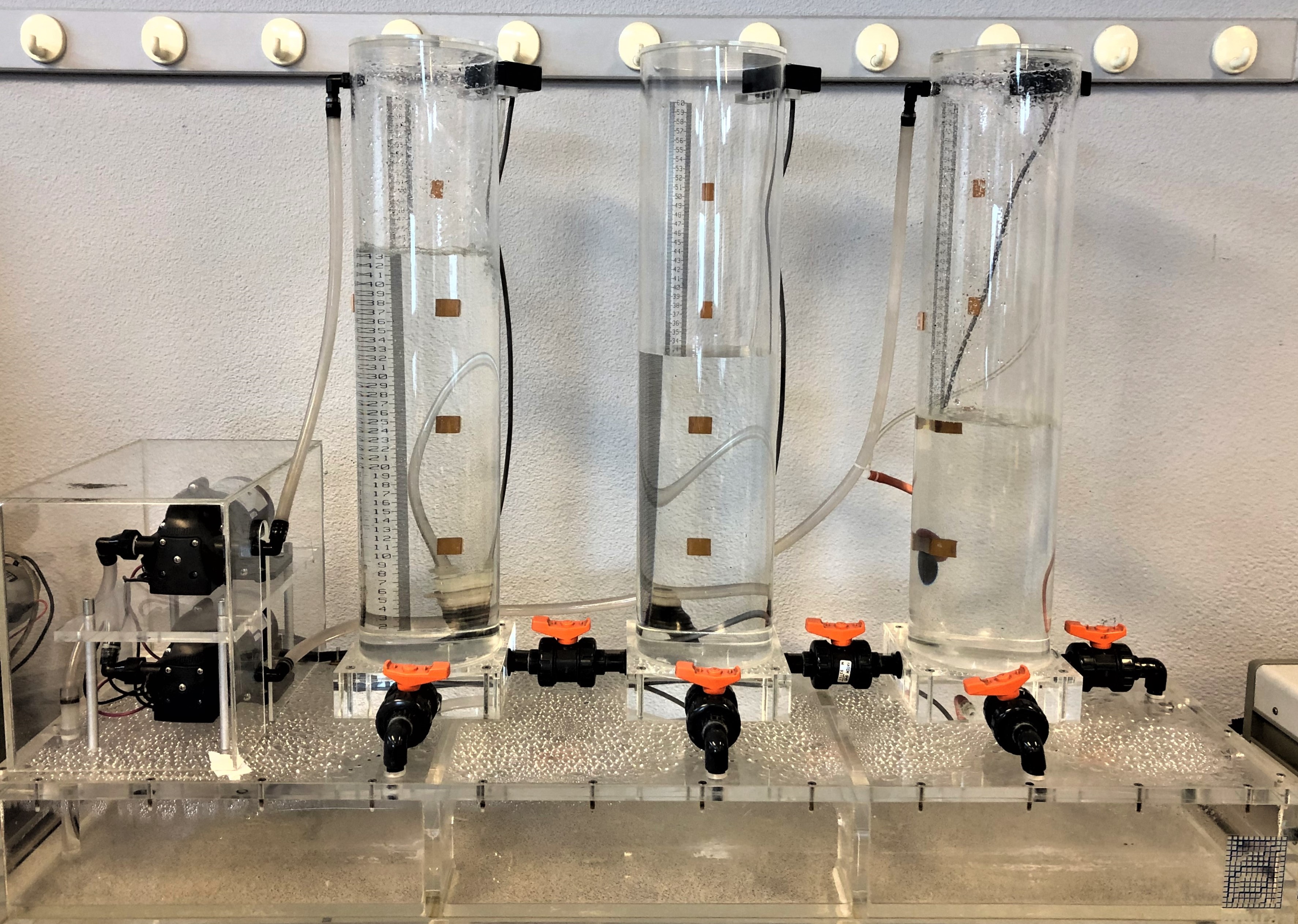}
\caption{\small Three-tank system.}
\label{Figure1}
\end{figure}

In this paper, all the experiments were performed on the three-tank system represented in Figure~\ref{Figure1}.

The three-tank system is about three cylindrical tanks of cross-sectional area $S$, connected to each other by means of pipes of section $S_n$. Each tank is also connected to the central reservoir through a pipe. Three piezo-resistive pressure transducer are installed on the top of each tank to measure the corresponding water level. The water is pumped from the central reservoir into tanks $T_1$ and $T_2$ with the help of two pumps (actuators) $P_1$ and $P_2$. The incoming flows, by unit of surface $S$, into tanks $T_1$ and $T_2$ are denoted by $u_1(t)$ and $u_2(t)$ and are obtained by using two electrical pumps with voltage input commands $u_{A_1}(t)$ and $u_{A_2}(t)$. The actuator dynamics are given by $\dot{u}_i(t)=-\frac{1}{T}u_i(t)+\frac{K}{T}u_{A_i}(t)$, $i=1,2$, where $T$ and $K$ denote, respectively, the time constant and the gain of each actuator's transfer function. Given that the actuator dynamics are negligible with respect to the system dynamics, in the sequel it is assumed that $u_i(t)=Ku_{A_i}(t)$. 

\begin{figure}[htb!]
\centering
\includegraphics[width=\columnwidth]{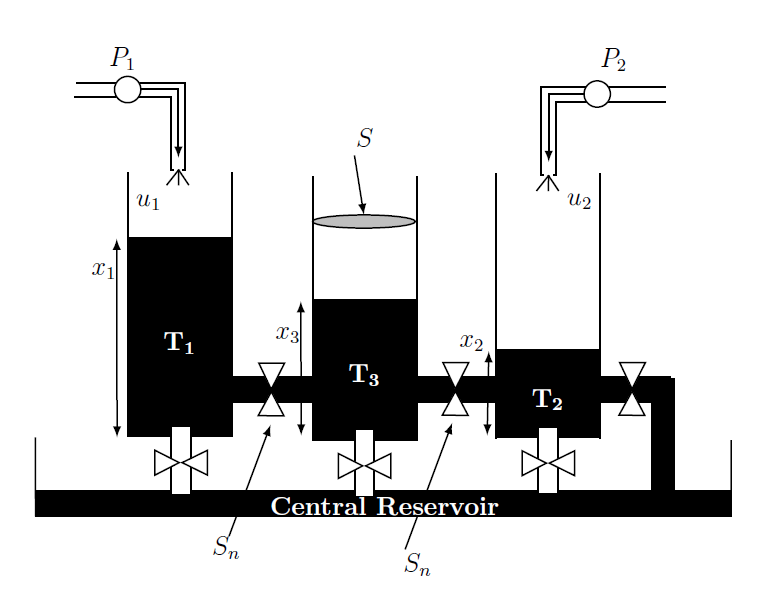}
\caption{\small Scheme of the three-tank System, Source: \cite{noura2009fault}}
\label{Figure2}
\end{figure}

The water level in the tank $T_i$ is denoted by $x_i(t)\geq 0$, $i=1,2,3$. The maximum water level in any tank is denoted by $h_{max}$ and the maximum incoming flow rate is denoted by $u_{max}$. A descriptive scheme of the system is presented in Figure~\ref{Figure2}.

The explicit system of equations that describes the dynamic of the three-tank model is given by:\footnote{In some of the following equations, the parentheses $(t)$ are dropped to save space.}
\begin{align}
\dot{x}_1 & = -Q_{10}(x_1)-Q_{13}(x_1,x_3)+u_1 \label{eq1}\\
\dot{x}_2 & = -Q_{20}(x_2)+Q_{32}(x_2,x_3)+u_2 \label{eq2} \\
\dot{x}_3 & = Q_{13}(x_1,x_3)-Q_{32}(x_2,x_3)-Q_{30}(x_3)\label{eq3}
\end{align}
where $Q_{i0}$, $i=1,2,3$ represents the outflow by unit of surface $S$ between each tank and the central reservoir, $Q_{13}$ is the outflow between tank $T_1$ and tank $T_3$ and $Q_{32}$ the outflow between tank $T_3$ and tank $T_2$. In this study, the valves that linking tanks $T_1$ and $T_3$ to the central reservoir are considered closed, \ie $Q_{10} \equiv 0$ and $Q_{30}\equiv 0$. The outflows by unit of surface $S$ are given by:
\begin{align}
Q_{13}(x_1,x_3) & = \mu_{13}\sgn(x_1-x_3)\sqrt{|x_1-x_3|} \label{q13}\\
Q_{20}(x_2) & = \mu_{20}\sgn(x_2)\sqrt{|x_2|} \label{q20}\\
Q_{32}(x_2,x_3) & = \mu_{32}\sgn(x_2-x_3)\sqrt{|x_3-x_2|} \label{q32}
\end{align}
where $\mu_{13}$, $\mu_{20}$ and $\mu_{32}$ are the flow coefficients and $\sgn$ is the sign function given by:
\begin{equation}
   \sgn(x)=\begin{cases}
   1 \quad \text{if} \quad x>0\\
   0 \quad \text{if} \quad x=0 \\
   -1 \quad \text{if} \quad x<0
   \end{cases}.
\end{equation}
The parameters' values of the three-tank system are given in Table~\ref{tb1}. 
\begin{table}[htb!]%
\begin{center}
\begin{tabular}{lcr}
\hline 
Parameters & Symbol &  Value \quad\quad \\ \hline
Tank sectional area & $S$ & $0.0154\; \SI{}{m^2}$ \quad \\ 
Pipes sectional area & $S_n$ & $5\times 10^{-5}\; \SI{}{m^2}$ \\ 
Outflow coefficient & $\mu_{13}$ & $8.5273\times 10^{-5}$\\ 
Outflow coefficient & $\mu_{32}$ & $8.5563\times 10^{-5}$\\ 
Outflow coefficient & $\mu_{20}$ & $1.5901\times 10^{-4}$ \quad \\ 
Maximum water level & $h_{max}$ & $0.62 \;\SI{}{m}\quad\quad$ \\ 
Maximum flow rate & $u_{max}$ & $10^{-4}\;\SI{}{m^3/s}\;\;$ \\ \hline
\end{tabular}
\caption{\small Parameters' values of the three-tank benchmark.}
\label{tb1}
\end{center}
\end{table}

The experimental setup is composed of a DTS200 three-tank benchmark connected to a Windows~7 PC using a MF624 plug-in card. Matlab/Simulink version R2008b is available on the development PC. 

For the FDI purpose, the three-tank model allows the application of any type of fault on both sensors and actuators:
\begin{itemize}
\item Multiplicative faults: sensor and actuator gains may be reduced from $100\%$ (total measurement) to $0\%$ (complete measurement failure);
\item Additive faults: sensors and actuators may present biases on their measurements.
\end{itemize}
Therefore, the sum of sensor and actuator faults can be expressed mathematically by \cite{rao2013modeling}:
\begin{align*}
\mathsf{S}_i^f(t)&=\alpha_i\mathsf{S}_i(t)+\mathsf{S}_{i0}\\
\mathsf{A}_j^f(t)&=\beta_j\mathsf{A}_j(t)+\mathsf{A}_{j0}
\end{align*}
where $\mathsf{S}_i^f(t)$ and $\mathsf{S}_i(t)$ (resp. $\mathsf{A}_j^f(t)$ and $\mathsf{A}_j(t)$) denote faulty and unfaulty $i$\textsuperscript{th} sensor (resp. $j$\textsuperscript{th} actuator) respectively, $\mathsf{S}_{i0}$ and $\mathsf{A}_{j0}$ are the biases (additive faults) of $i$\textsuperscript{th} sensor and $j$\textsuperscript{th} actuator respectively, and $0\leq \alpha_i \leq1$ and $0\leq\beta_j\leq1$ are gain loss factors (multiplicative faults). The method presented in this paper can be used for both multiplicative and additive faults.

\textit{Hypothesis}: In the sequel, we assume that there is only one fault at a time affecting sensors or actuators. 

In the next section, we introduce a new definition of the signature matrix by taking into account the sensitivity of the residues with respect to the faults.

\section{Flatness-Based FDI}
\label{section3}
\subsection{Flatness-based residual generation}
Consider the following nonlinear system
\begin{equation}
\begin{cases}
\dot{x}=f(x,u)\\
y=h(x,u)
\label{nonlinear}
\end{cases}
\end{equation}     
where $x=(x_1,\ldots,x_n)^T$ is the state vector, belongs to an n-dimensional manifold $X$, $u=(u_1,\ldots,u_m)^T \in \mathbb{R}^m$ is the input vector, $y=(y_1,\ldots,y_p)^T \in \mathbb{R}^p$ is the measured output, $m\leq n$, $p \geq m$ and $\rang\big(\frac{\partial f}{\partial u}\big)=m$. In the sequel, we denote by
\begin{align}
    \overline{\xi}=(\xi,\dot{\xi},\ddot{\xi},\ldots) \in \mathbb{R}_\infty^m
\end{align}the sequence of infinite order jets of a vector $\xi$ and by
\begin{align}
    \overline{\xi}^{(\alpha)}\triangleq(\xi,\dot{\xi},\ldots,\xi^{(\alpha)})
    \label{truncation_var}
\end{align}the truncation at the finite order $\alpha \in \mathbb{N}$. Let $(x,\overline{u})\triangleq (x,u,\dot{u},\ddot{u},\ldots)$ be a prolongation of the coordinates $(x,u)$ to the manifold of jets of infinite order $\mathfrak{X} \triangleq X \times \mathbb{R}_\infty^m$ \cite[Chapter 5]{levine2009analysis}. 
\begin{defi}[\cite{fliess1993differentially}]\label{flatness}
The system \eqref{nonlinear} is said to be flat at a point $(x_0,\overline{u}_0) \in \mathfrak{X}$ if, and only if, there exist a vector $z=(z_1,\ldots, z_m)^T \in \mathbb{R}^m$ and two mappings $\psi$, defined on a neighbourhood $\mathcal{V}$ of $(x_0,\overline{u}_0) \in \mathfrak{X}$, and $\varphi=(\varphi_0,\varphi_1,\ldots)$, defined on a neighbourhood $\mathcal{W}$ of $\psi (\mathcal{V})$ of $\overline{z}=(z,\dot{z},\ldots)\triangleq \psi (x_0,\overline{u}_0)$ such that:
\begin{enumerate}
\item $z$ is a function of $x$, $u$ and successive derivatives of $u$ up to a finite order $\nu$:
\begin{equation}
z=\psi(x,u,\dot{u},\ldots,u^{(\nu)});
\label{z}
\end{equation}
\item In turn, $x$ and $u$ are functions of $z$ and its successive derivatives up to a finite order $\rho$:
\begin{equation}
(x,u)=\\(\varphi_0(z,\dot{z},\ldots,z^{(\rho)}),\varphi_1(z,\dot{z},\ldots,z^{(\rho+1)})),
\label{x,u}
\end{equation}
hence, the expression of the output $y$ is given by:
\begin{equation}
   y= h(\varphi_0(z,\dot{z},\ldots,z^{(\rho)}),\varphi_1(z,\dot{z},\ldots,z^{(\rho+1)}));
   \label{y}
\end{equation}
\item The differential equation $\frac{d\varphi_0}{dt} = f(\varphi_{0}, \varphi_{1})$ is identically satisfied.
\end{enumerate}
 The vector $z$ is called flat output of the system and its components $z_1,\ldots,z_m$ and their successive derivatives are linearly independent. The mappings $\psi$ and $\varphi$ are called isomorphisms of Lie-B\"{a}cklund and are inverse of one another. 
\end{defi}

\begin{rem}[\cite{kaminski2018intrinsic}] The property of flatness is not defined globally on the state space $\mathfrak{X}$. It means that there may exist points on $\mathfrak{X}$ where the system is not flat, or, in other words, where the isomorphisms of Lie-B\"{a}cklund $\psi$ and $\varphi$ are not defined. The set of such points is called the set of intrinsic singularities. In \cite{kaminski2018intrinsic} it is shown that the set of equilibrium points that are not first order controllable, is included in the set of intrinsic singularities.
\label{intrinsic}
\end{rem}

Let us assume the system \eqref{nonlinear} is flat with $z=(z_1,\ldots,z_m)^T$ as flat output. We also suppose that the full output $y$ is measured by sensors $\mathsf{S}_1,\ldots,\mathsf{S}_p$ and we denote its measurement by 
\begin{equation}
y^s=(y_1^s,\ldots,y_p^s)^T.
\label{output}
\end{equation}
Moreover, we assume that the values $u_1,\ldots,u_m$ of the input vector $u$, corresponding to the actuators $\mathsf{A}_1,\ldots,\mathsf{A}_m$, are available at any moment.

In order to detect and isolate faults on physical sensors and actuators, their analytical measurements must first be computed. Equations \eqref{x,u} and \eqref{y} provide an efficient way to construct these analytical measurements, as long as the measurement of the flat output is available during the system process. In the following, we suppose that the measurement of the flat output is available and we denote it by:
\begin{equation}
    z^s=(z_1^s,\ldots,z_m^s)^T.
    \label{measured_fo}
\end{equation}
According to \eqref{x,u}, the analytical state $x^z$ and input $u^z$, constructed via the flat output \eqref{measured_fo}, read:
\begin{equation}
x^z=\varphi_0(\overline{z^s}^{(\rho)}) \quad \mathrm{and} \quad u^z=\varphi_1(\overline{z^s}^{(\rho+1)}),
\label{xz,uz}
\end{equation}
and the analytical output $y_k^z$, is given, according to \eqref{y}, by:
\begin{equation}
y_k^z\triangleq h_k(\varphi_0(\overline{z^s}^{(\rho)}),\varphi_1(\overline{z^s}^{(\rho +1)})).
\label{virtual-output}
\end{equation}
The variables $\overline{z^s}^{(\rho)}$ and $\overline{z^s}^{(\rho +1)}$ are defined using \eqref{truncation_var}.

The following definition of residues is borrowed from \cite{rammal2020choice}:
\begin{defi}
The k\textsuperscript{th}-sensor residue $R_{\mathsf{S}_k}$, for $k=1, \ldots, p$, and the l\textsuperscript{th}-input residue $R_{\mathsf{A}_l}$, for $l=1, \ldots, m$, are given by:
\begin{align}
 R_{\mathsf{S}_k}= y_k^s - y_k^z,  \quad R_{\mathsf{A}_l}=u_l-u_l^z,
 \label{residue}
 \end{align}
respectively.
\end{defi} 
Then, the full vector of residues, denoted by $r$, is of dimension $p+m$ and given by:
\begin{align}
r&=(R_{\mathsf{S}_1},\ldots,R_{\mathsf{S}_m},R_{\mathsf{S}_{m+1}},\ldots,R_{\mathsf{S}_p},R_{\mathsf{A}_1},\ldots,R_{\mathsf{A}_m})^T\nonumber \\&=(r_1,\ldots,r_m,r_{m+1},\ldots, r_p, r_{p+1},\ldots,r_{p+m})^T.
\label{full-residue}
\end{align}

\begin{rem}\label{rem1}
We can consider, without loss of generality, that:
\begin{equation}
    z^s=(y_1^s,\ldots,y_m^s)^T.
    \label{particular_zs}
\end{equation}
In this case, the first $m$ components of $y^z$ are equal to the corresponding components of $z^s$, then
\begin{align}
y^z=(z^s,\widetilde{h}(\varphi_0(\overline{z^s}^{(\rho)}),\varphi_1(\overline{z^s}^{(\rho +1)})))^T
\label{yz}
\end{align}
with 
\begin{align*}
\widetilde{h}=h_{m+1}(\varphi_0(\overline{z^s}^{(\rho)}),&\varphi_1(\overline{z^s}^{(\rho +1)}),\ldots,\\ & h_{p}(\varphi_0(\overline{z^s}^{(\rho)}),\varphi_1(\overline{z^s}^{(\rho +1)}).
\end{align*}
Hence, the first $m$ residues are identically zero, and the vector \eqref{full-residue} becomes:
\begin{align}
r=(0,\ldots,0,r_{m+1},\ldots, r_p, r_{p+1},\ldots,r_{p+m})^T.
\label{zero-residue}
\end{align}

A zero residue means that even if a fault occurs on one sensor or actuator, this residue cannot be affected. Then, it is not useful either for detection or isolation of the fault and we eliminate it from \eqref{zero-residue}, which will be truncated to the last $p$ components:
\begin{align}
r_\tau&=(R_{\mathsf{S}_{m+1}},\ldots,R_{\mathsf{S}_p},R_{\mathsf{A}_1},\ldots,R_{\mathsf{A}_m})^T\nonumber \\&=(r_{\tau_1},r_{\tau_2},\ldots,r_{\tau_p})^T.
\label{truncated-residue}
\end{align} 
\end{rem}

\begin{rem}
The components $z_1^s,\ldots,z_m^s$ of the flat output \eqref{measured_fo}, must be derivated in order to calculate the values of $y^z$ and $u^z$. Due to the presence of noise on system's sensors and actuators, filtering these derivatives is inevitable. Many methods have been developed in the literature and can be used here, we cite among them the algebraic derivative estimation \cite{zehetner2007derivative}, high-gain observers \cite{vasiljevic2008error} and averaged finite difference methods \cite{anderssen1998stable}. %
In our experiments on the real three-tank system, we use the algebraic derivative estimation based on a receding horizon strategy given in \cite{zehetner2007derivative}:
\begin{equation}
    y^{(j)}_k \approx (-1)^{(j)}\frac{Ts}{2}\sum_{i=1}^M(\Pi_{i-1}y_{k-i+1}+\Pi_{i}\, y_{k-i})
\end{equation}
where $y^{(j)}_k=y^{(j)}(kTs)$ is the $j\textsuperscript{th}$ time derivative of $y(kTs)$, $y_{k-i}=y(t_{k-i})$ with $t_{k-i}=(k-i)Ts$ and $T_s=1 \SI{}{s}$ is the sample time, $M=T/Ts$ with $T$ is small time window and $\Pi_i=\Pi_{jN\nu}(T,ti)$ with
\begin{align}
   \Pi_{j N\nu}&( T,t_i)=\frac{(N+j+\nu +1)!(N+1)!(-1)^j}{T^{N+j+\nu +1}} \nonumber\\
 & \sum_{\kappa_1=0}^{N-j}\sum_{\kappa_2 =0}^j\Big ( \frac{(T-t_i)^{\nu+\kappa_1+\kappa_2}(-t_i)^{N-\kappa_1-\kappa_2}}{\kappa_1 !\kappa_2 !(N-j-\kappa_1)!(j-\kappa_2)!}\nonumber\\ &\frac{1}{(N-\kappa_1-\kappa_2)!(\nu+\kappa_1+\kappa_2)!(N-\kappa_1+1)}\Big ) 
\end{align} with $N$ the order of the Taylor-series expansion and $\nu$ is the number of additional integrals.
\end{rem} 

\subsection{Fault Detection and Isolation}
In order to detect and isolate faults using the flatness-based approach, a definition of the signature matrix was introduced in \cite{rammal2020choice}. However, the former definition does not take into account the sensitivity of the residues with respect to the faults. Indeed, sometimes even if the residue depends on the faulty signal, it is not sensitive enough to exceed its threshold. Therefore, in this paper we propose a new definition of the signature matrix definition:
\begin{defi}[Signature matrix]\label{def_signature}
Given the vector of residues $r$ defined in \eqref{truncated-residue} and $\zeta=(y_{1}^s,\ldots,y_{p}^s,u_{1},\ldots,u_{m})$ the vector of measurements that are subject to faults, the signature matrix $\mathbf{S}$, associated to $z^s$, is given by:  
\begin{equation}
\mathbf{S}=\begin{pmatrix}
\sigma_{1,1} &  \sigma_{1,2} & \ldots & \sigma_{1,p+m} \\
\vdots & \vdots & \ldots & \vdots \\
\sigma_{p,1} & \sigma_{p,2} & \ldots & \sigma_{p,p+m}
\end{pmatrix}
\label{signature-matrix}
\end{equation}
with
\begin{equation} \label{sigma}
    \sigma_{i,j}\triangleq
    \begin{cases}
    0\; \text{if}\; \frac{\partial r_{\tau_i}}{\partial \zeta_j^{(\varrho)}}=0 \quad \forall \varrho \in \lbrace 0,1,\ldots, \rho +1\rbrace,\\
    1\; \text{if}\; \exists\; \varrho \in \lbrace 0,\ldots,\rho +1\rbrace\text{ s.t. } \Big\lvert \frac{\partial r_{\tau_i}}{\partial \zeta_j^{(\varrho)}} \Big\rvert > \mathbf{Th}_{i,j,\varrho}
    \end{cases}
\end{equation}
where $\mathbf{Th}_{i,j,\varrho}$ is the threshold of the sensitivity of the residue $r_{\tau_i}$ with respect to the $\varrho$-order derivative of the variable $\zeta_j$.
The partial derivation is defined with respect to the coordinates~$(x,u)$ and their prolongation.
\end{defi} 
Let $\Sigma_j$, for $j=1,\ldots,p+m$, be the j\textsuperscript{th}-column of the matrix $\mathbf{S}$. $\Sigma_j$ indicates if a residue $r_{i}$ is or is not functionally affected by a fault on the measurement $\zeta_j$: $\sigma_{i,j}=0$ means that $r_{i}$ is not affected by a fault on $\zeta_j$ and $\sigma_{i,j}=1$ if it is affected. $\mathbf{Th}_{i,j,\varrho}$ can be found experimentally if sufficient experimental data with and without faults is available or theoretically by considering also the relative influence of disturbances and uncertainties on the residue  (see also  \cite[Chapter~7]{ding2008model}).

\begin{defi}[Fault alarm signature]
Each column $\Sigma_j$ of the signature matrix $\mathbf{S}$ is called fault alarm signature or simply signature, associated to the sensor/actuator $\zeta_j$.
\end{defi} 

\begin{rem}
Given that the flat outputs are measured by sensors (see Remark~\ref{rem1}), the dimension of the signature matrix $\mathbf{S}$ will be reduced to $p\times (p+m)$.
\end{rem}
The following definitions of \emph{detectability} and \emph{isolability} in the flatness context are borrowed from \cite{rammal2020choice}:

\begin{defi}[Detectability]\label{detectability}
 A fault on a sensor/actuator $\zeta_j$ is detectable if there exists at least one $i \in\lbrace 1,\ldots,p+m\rbrace$ such that $\sigma_{i,j}=1$.
\end{defi} 

\begin{defi}[Isolability]\label{isolability}
 A fault on a sensor $\mathsf{S}_k$, \\ $k=1,\ldots,p$, is said isolable if its corresponding fault alarm signature $\Sigma_k$ in the signature matrix $\mathbf{S}$ is distinct from the others, \ie 
\begin{equation}
\Sigma_k\neq \Sigma_j, \quad \forall j=1,\ldots,p+m, \; j\neq k .
\label{isol_sigma}
\end{equation} 
An isolable fault on the actuator $\mathsf{A}_l$, for $l=1,\ldots,m$, is defined analogously: 
\begin{equation}
\Sigma_{p+l}\neq \Sigma_j, \quad \forall j=1,\ldots,p+m, \; j\neq p+l.
\label{13-1}
\end{equation}
\end{defi}

This definition of isolability reflects the fact that if the signature matrix $\mathbf{S}$ has two identical signatures $\Sigma_i=\Sigma_j$ with $i\neq j$, then a fault that affects the sensor/actuator $\zeta_i$ or $\zeta_j$ cannot be isolated. Therefore, in order to be able to isolate as many faults as possible, we need to increase the number of the residues by using several flat outputs. These flat outputs must be independent, in the sense that if a fault affects one flat output, not all the residues will be affected \cite{torres2013fault}. In the next section, a characterization of the relation between flat outputs is discussed.

\section{Flat Output Characterization}\label{section4}
We suppose that the flat system \eqref{nonlinear} admits different flat outputs whose measurements are available. We also define by $\mu$ the number of distinct signatures of the matrix $\mathbf{S}$, associated to a flat output, then $\mu$ is the number of isolated faults. So, in order to get more isolability of faults, we need to increase the number of distinct signatures $\mu$.

In the following, we denote the i\textsuperscript{th} element of the set of $q$ flat output vectors $Z_i$ by
\begin{equation}
Z_i=(z_{i1},\ldots,z_{im})^T.
\label{notation}
\end{equation}
In order to characterize the flat outputs, the notion of augmented signature matrix is defined:
\begin{defi}[Augmented signature matrix]
Let $Z_1,\ldots, Z_q$ be $q$ different measured flat outputs of the flat system \eqref{nonlinear}. The augmented signature matrix $\widetilde{\mathbf{S}}$ associated to $Z_1,\ldots,Z_q$ is defined by:
\begin{equation}
\widetilde{\mathbf{S}}=\begin{pmatrix}
\mathbf{S}_1 \\ \mathbf{S}_2 \\ \vdots \\ \mathbf{S}_q
\end{pmatrix}
\label{25}
\end{equation}
where $\mathbf{S}_i$ is the signature matrix associated to the flat output vector $Z_i$.
\end{defi}

\begin{defi}[Independence]\label{def_independence}
Let $\widetilde{\mathbf{S}}$ be the augmented signature matrix associated to $Z_1$ and $Z_2$:
\begin{equation*}
\widetilde{\mathbf{S}}=\begin{pmatrix}
\mathbf{S}_1 \\ \mathbf{S}_2
\end{pmatrix},
\end{equation*}
$\mu_i$, $i=1,2$, the number of distinct signatures of the matrix $\mathbf{S}_i$ and $\widetilde{\mu}$ the number of distinct signatures of the augmented matrix $\widetilde{\mathbf{S}}$. We say that $Z_1$ and $Z_2$ are independent if, and only if, 
\begin{equation}
\widetilde{\mu}> \mu_1 \quad \text{and} \quad \widetilde{\mu}> \mu_2 .
\label{condition}
\end{equation}
\end{defi}

According to Definition~\ref{def_independence}, the condition of full isolability is achieved if the augmented matrix 
\begin{align}
\widetilde{\mathbf{S}}=\begin{pmatrix}
\mathbf{S}_1 \\ \mathbf{S}_2 \\ \vdots \\ \mathbf{S}_q
\end{pmatrix}
\end{align}
has $p+m$ distinct signatures, \ie $\widetilde{\mu}=p+m$.

In \cite{rammal2020choice}, this characterization of the flat outputs is applied on the three-tank system in the open loop case, and two flat outputs are needed to achieve full isolability. In the open loop case, a fault that affects a sensor has no impact on the actuators. 

In this paper, we focus on the closed-loop case. As such, the control action depends on the measured outputs. Therefore, a fault in a closed-loop system is propagated with the feedback loop, increasing the difficulty of fault isolation. For the system \eqref{nonlinear}, this implies that at least one input component $u_l$ is related to an output measurement $y_k^s$. Then, the expression of the control input $u_l$ is given by:
\begin{equation}
    u_l=u_l^{ref}+C_{l,k}(y_k^{ref}-y_k^s)
    \label{closed-loop}
\end{equation}
where, $u_l^{ref}$ and $y_k^{ref}$ are the reference input and output trajectories, respectively, and $C_{l,k}$ denotes the discrete-time feedback controller between the $l$\textsuperscript{th} control input and the $k$\textsuperscript{th} measured output. In this case, the $l$\textsuperscript{th} input residue $R_{\mathsf{A}_l}$ is given by:
\begin{equation}
    R_{\mathsf{A}_l}=u_l-u_l^z
\end{equation}
where $u_l$ is replaced by \eqref{closed-loop}. So a fault that appears on the sensor $\mathsf{S}_k$ and affects originally only the residue $R_{\mathsf{S}_k}$, in the closed-loop case, affects both residues $R_{\mathsf{S}_k}$ and $R_{\mathsf{A}_l}$.

\section{Experimental Results}
\label{section5}
In this section, experiments on the three-tank system are presented in order to show the effectiveness of the flatness-based FDI method and the characterization of the flat outputs. First, subsection~\ref{sec5.1} represents the flatness analysis of the three-tank model. Then, subsection~\ref{section5.2} represents the path tracking, using the flatness property, and the the controller applied on the system. Finally, subsection~\ref{sec5.3} represents theoretical and experimental results of FDI.

\subsection{Flatness analysis of the three-tank system}\label{sec5.1}
In the three-tank system, the only equilibrium point which is not first order controllable, or that represents an intrinsic singularity (see Remark~\ref{intrinsic}) is where the water level is equal in all three tanks, \ie $x_1=x_2=x_3$. To avoid this singularity, we consider the following configuration:
\begin{align}
(\mathcal{C}):\quad x_1>x_3>x_2>0. 
\end{align}
According to Definition~\ref{flatness}, we can show that the three-tank system \eqref{eq1}-\eqref{eq2}-\eqref{eq3} is flat with
\begin{align}
z=(x_1,x_3)^T=(z_1,z_2)^T
\label{8}
\end{align} a flat output. In fact, from \eqref{eq3} and using \eqref{q13} and \eqref{q32}, it is easy to express $x_2$ in function of $z$:
\begin{align}
x_2=z_2-\frac{1}{\mu_{32}^2}\Big(-\dot{z}_2+\mu_{13}\sqrt{z_1-z_2}\Big)^2.
\label{x2}
\end{align}
In addition, from \eqref{eq1} and \eqref{q13}, $u_1$ is given by:
\begin{align}
u_1=\dot{z}_1+\mu_{13}\sqrt{z_1-z_2}\; ,
\label{u1z1}
\end{align}
and from \eqref{eq2}, $u_2$ is expressed by
\begin{align}
u_2=\dot{x}_2+\mu_{20}\sqrt{x_2}-\mu_{32}\sqrt{z_2-x_2}
\label{u2z1}
\end{align}
where $x_2$ is given in \eqref{x2} as a function of $(x_1,x_3)$, which proves that $z=(x_1,x_3)^T$ is a flat output of the three-tank system.

\subsection{Path tracking and control of the system}
\label{section5.2}
In control theory, the concept of path tracking consists in finding control input values allowing the system to follow a predefined reference trajectory. The flatness property ensures the calculation of the control variables in a very simple way. One needs only to calculate a trajectory $t\mapsto z_{ref}(t)$, sufficiently differentiable, for the flat output $z$ then, since the variables $x$ and $u$ are functions of $z$ and its successive derivatives, the reference trajectory $t\mapsto (x_{ref}(t),u_{ref}(t))$ of the system is deduced by differentiating $t\mapsto z_{ref}(t)$ a finite number of times. Concerning our three-tank system experiments, a reference trajectory of the system is generated using the flat output $z=(x_1,x_3)^T$. The initial and final conditions for the flat output are given by:
\begin{align}
x_{1_i}= 0.20\; \SI{}{m},\quad x_{1_f}=0.35\; \SI{}{m} \nonumber\\
x_{3_i}= 0.15 \;\SI{}{m},\quad x_{3_f}=0.25\; \SI{}{m}\label{12}
\end{align}
The initial and final times are respectively $t_i=0\, \SI{}{s}$ and $t_f=400 \,\SI{}{s}$.  The trajectory $t\mapsto z_{ref}(t)$, that does not satisfy any differential equation, is calculated using a fifth order polynomial interpolation \cite{levine2009analysis}.

\begin{figure}[htb!]
\centering
\includegraphics[width=\columnwidth]{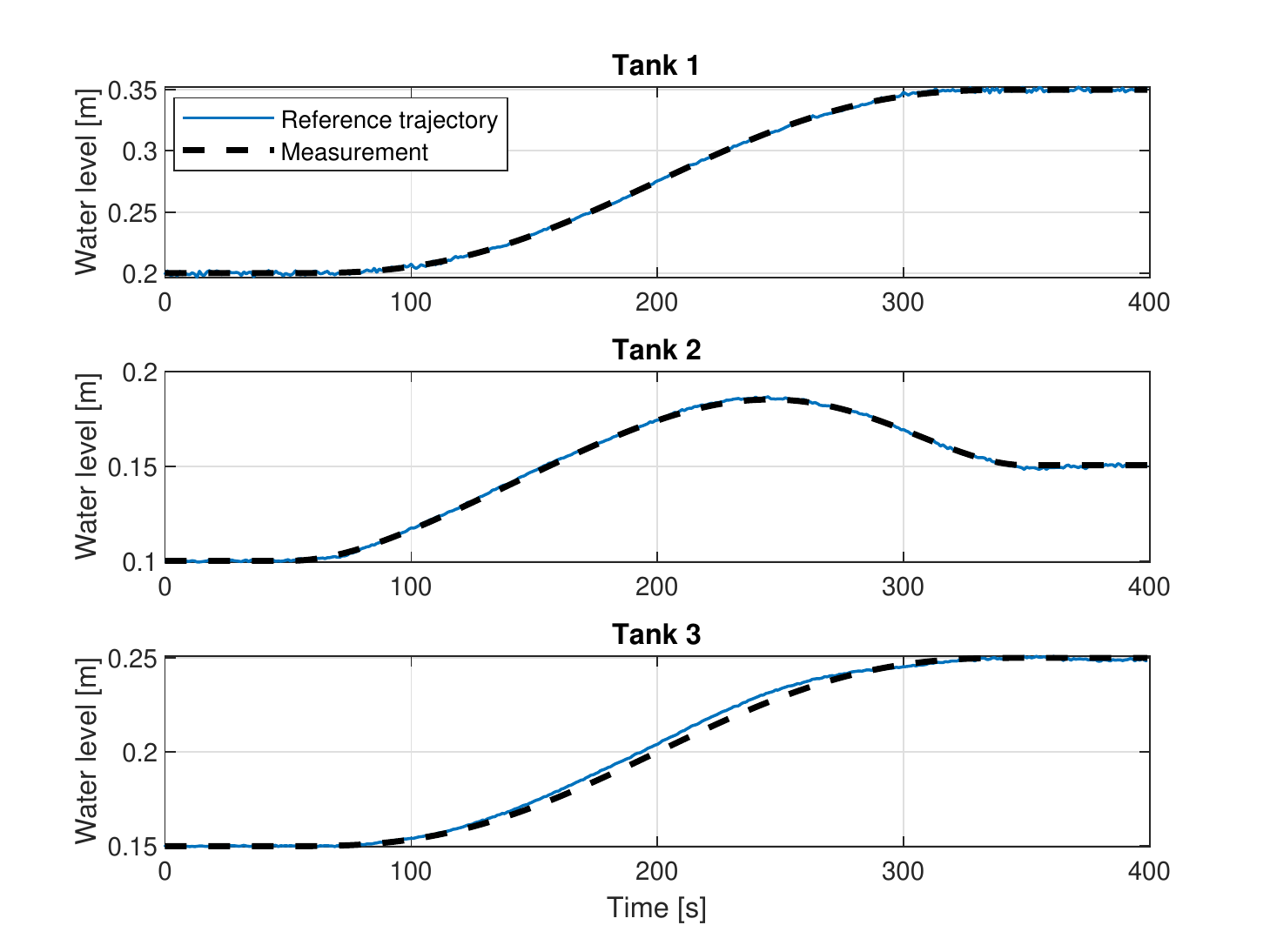}
\caption{\textit{Reference trajectories vs. measurements of the water level in each tank, in the fault free case.}}
\label{Figure3}
\end{figure}

In order to control the system against external faults and disturbances, the following PI controllers, that are already implemented on the system, are used to control the water level in tanks $T_1$ and $T_2$ \cite{martinez2014flatness}:
\begin{align}
    C_{1,1}(z)&=\frac{-0.001043+0.0009565z^{-1}}{1 -z^{-1}}\\
    C_{2,2}(z)&=\frac{-0.00104+0.00096z^{-1}}{1 -z^{-1}}
\end{align}The output of the system in the fault free case is represented in Figure~\ref{Figure3}.

\subsection{Results on FDI}
\label{sec5.3}
Sensors on the three-tank system are $\mathsf{S}_1$, $\mathsf{S}_2$ and $\mathsf{S}_3$ and they measure the water level in each tank, respectively so the measured output is given by:
\begin{equation}
y^s=(x_1^s,x_2^s,x_3^s)^T \triangleq (y_1^s,y_2^s,y_3^s)^T .
\end{equation}
Pumps $P_1$ and $P_2$ are the actuators of the system and we denote them by $\mathsf{A}_1$ and $\mathsf{A}_2$ and their outgoing flows are denoted by $u_1$ and $u_2$, respectively. 

In this paper, we show that a single flat output vector is not sufficient to isolate all possible faults on the three-tank system and that we need a second flat output to ensure full isolability. So two cases are represented: using one flat output and using two flat outputs.

\subsubsection*{Case A: using one flat output}
As shown in section~\ref{sec5.1}, the three-tank system is flat with $z=(x_1,x_3)^T$ as flat output. Components of the flat output are measured by sensors $\mathsf{S}_1$ and $\mathsf{S}_3$ and their measurements are denoted by:
\begin{equation}
z^s=(y_1^s,y_3^s)^T\triangleq (z_1^s,z_2^s)^T .
\label{flat1}
\end{equation}
In order to construct the vector of residues, the redundant inputs and outputs are firstly computed using \eqref{8} through \eqref{u2z1}:
\begin{align}\label{redundant}
y_1^z&=z_{1}^s \nonumber \\
y_2^z&=z_{2}^s-\frac{1}{\mu_{32}^2}\Big(-\dot{z_2^s}+\mu_{13}\sqrt{z_1^s-z_2^s}\Big)^2 \nonumber \\
y_3^z & =z_{2}^s \\
u_1^z& =\dot{z}_{1}^s+\mu_{13}\sqrt{z_{1}^s-z_{2}^s} \nonumber \\
u_2^z &=\dot{y}_2^z-\mu_{32}\sqrt{z_{2}^s-y_2^z}+\mu_{20}\sqrt{ y_2^z}.\nonumber
\end{align}
Then, the vector of residues associated to $z^s$ is given by:
\begin{align}
r=\begin{pmatrix}
R_{\mathsf{S}_1}\\ R_{\mathsf{S}_2}\\R_{\mathsf{S}_3} \\  R_{\mathsf{A}_1} \\ R_{\mathsf{A}_2} 
\end{pmatrix}=\begin{pmatrix}
y_1^s\\ y_{2}^s\\y_3^s \\  u_{1} \\ u_{2} 
\end{pmatrix}- \begin{pmatrix}
 y_1^z\\ y_2^z\\ y_3^z \\  u_1^z \\ u_2^z
\end{pmatrix}.
\label{r1}
\end{align}
Nevertheless, according to Remark~\ref{rem1}, residues $R_{\mathsf{S}_1}$ and $R_{\mathsf{S}_3}$ are identically zero, hence the vector $r$ is truncated to:
 \begin{align}
r_\tau=( R_{\mathsf{S}_2},R_{\mathsf{A}_1},R_{\mathsf{A}_2})^T\triangleq (r_{\tau_1},r_{\tau_2},r_{\tau_3})^T.
\label{rtau1}
\end{align}
The vector $\zeta$ of measurements that are subject to faults, introduced in Definition~\ref{def_signature}, is given by:
\begin{equation}
\zeta=(y_1^s,y_2^s,y_3^s,u_1,u_2) \in \mathbb{R}^5.
\end{equation}
Therefore, the signature matrix $\mathbf{S}$, associated to $z^s$, is of dimension $3\times 5$:
\begin{equation}
\mathbf{S}=
\begin{pmatrix}
\sigma_{1,1} &  \sigma_{1,2} & \sigma_{1,3} & \sigma_{1,4} &\sigma_{1,5} \\
 \sigma_{2,1} & \sigma_{2,2} & \sigma_{2,3} & \sigma_{2,4} & \sigma_{2,5} \\
 \sigma_{3,1} & \sigma_{3,2} & \sigma_{3,3} & \sigma_{3,4} & \sigma_{3,5}
\end{pmatrix}
\label{sig}
\end{equation} 

\begin{table*}[htb!]
    \centering
    \begin{tabular}{|c|c|c|c|c|c|c|c|c|c|}
    \hline
         & $\pdv{y_1^s}$ & $\pdv{y_2^s}$ & $\pdv{y_3^s}$ & $\pdv{\dot{y}_1^s}$ & $\pdv{\dot{y}_2^s}$& $\pdv{\dot{y}_3^s}$ & $\pdv{\ddot{y}_3^s}$ & $\pdv{u_1}$ & $\pdv{u_2}$ \\ \hline
        $R_{S_2}$ & $ 0.99$ & 1 & $1.99$ & 0 & 0 & $5209$ & 0 & 0 & 0  \\ \hline
        $R_{A_1}$ &$10^{-4}$& 0 & $10^{-4}$ & 1 & 0 & 0 & 0 & 1 & 0 \\ \hline
        $R_{A_2}$ & $4\times 10^{-4}$ & 1 & $7\times 10^{-4}$ & $0.99$ & 0 & $4.3$ & 5209 & 0 & 1  \\ \hline
    \end{tabular}
    \caption{Partial derivatives of residues $R_{S_2}$, $R_{A_1}$, and $R_{A_2}$ with respect to $\zeta$ and its derivatives computed at the equilibrium point $x_{1_e}=0.20$~m, $x_{2_e}=0.10$~m, $x_{3_e}=0.15$~m.}
    \label{table_sens_Z1}
\end{table*}

According to Definition~\ref{def_signature}, in order to construct the signature matrix $\mathbf{S}$, a threshold has to be fixed for the sensitivity of each residue with respect to a measurement and its time derivatives. An analysis of the sensitivity can be made by computing the numerical values of the partial derivatives of the residues with respect to the measurements and their time derivatives around an equilibrium point. Table~\ref{table_sens_Z1} gives numerical values for the partial derivatives at the equilibrium state $x_{1_e}=0.20$~m, $x_{2_e}=0.10$~m, $x_{3_e}=0.15$~m. Higher absolute values with respect to measurements of sensors or actuators denote higher sensitivity. On the other hand, a small absolute value indicates that the residue can be insensitive to the fault in the presence of disturbances or uncertainties. Given the values from Table~\ref{table_sens_Z1}, a common threshold value $\mathbf{Th}=0.5$ for the sensitivity matrix is selected such that the residues with sensitivities of order $10^{-4}$ are neglected for the FDI analysis. It can be noticed that residue $R_{A_1}^{Z_1}$ is weakly sensitive to a fault on sensor $\mathsf{S}_3$ because its partial derivatives with respect to $y_3^s$ and its derivatives are smaller than $\mathbf{Th}$. The following signature matrix is obtained:
\begin{equation}
\mathbf{S}=\begin{pmatrix}
1 & 1 & 1 & 0& 0\\
1 & 0 & 0 & 1 & 0\\
1 & 1 & 1 & 0 & 1 
\end{pmatrix}.
\label{sig_z1_2}
\end{equation}

As mentioned in Section~\ref{section2}, both additive and multiplicative faults can be applied on the DTS200 three-tank system. In this paper, we add multiplicative faults on all the system sensors and actuators. Moreover, due to the presence of noise on the system, thresholds for each residue are fixed: several nominal experiments were run, \ie without introducing any fault on system sensors and actuators. For each experiment, the initial and final values of the reference trajectory are modified. The maximum and minimum values of the residues are extracted in each experiment, and the amplitude of the threshold is fixed by choosing the worst case among all the calculated residues. A safety margin of $5\%$ is added to avoid false alarms.

\begin{figure}[htb!]
\centering
\includegraphics[width=\columnwidth]{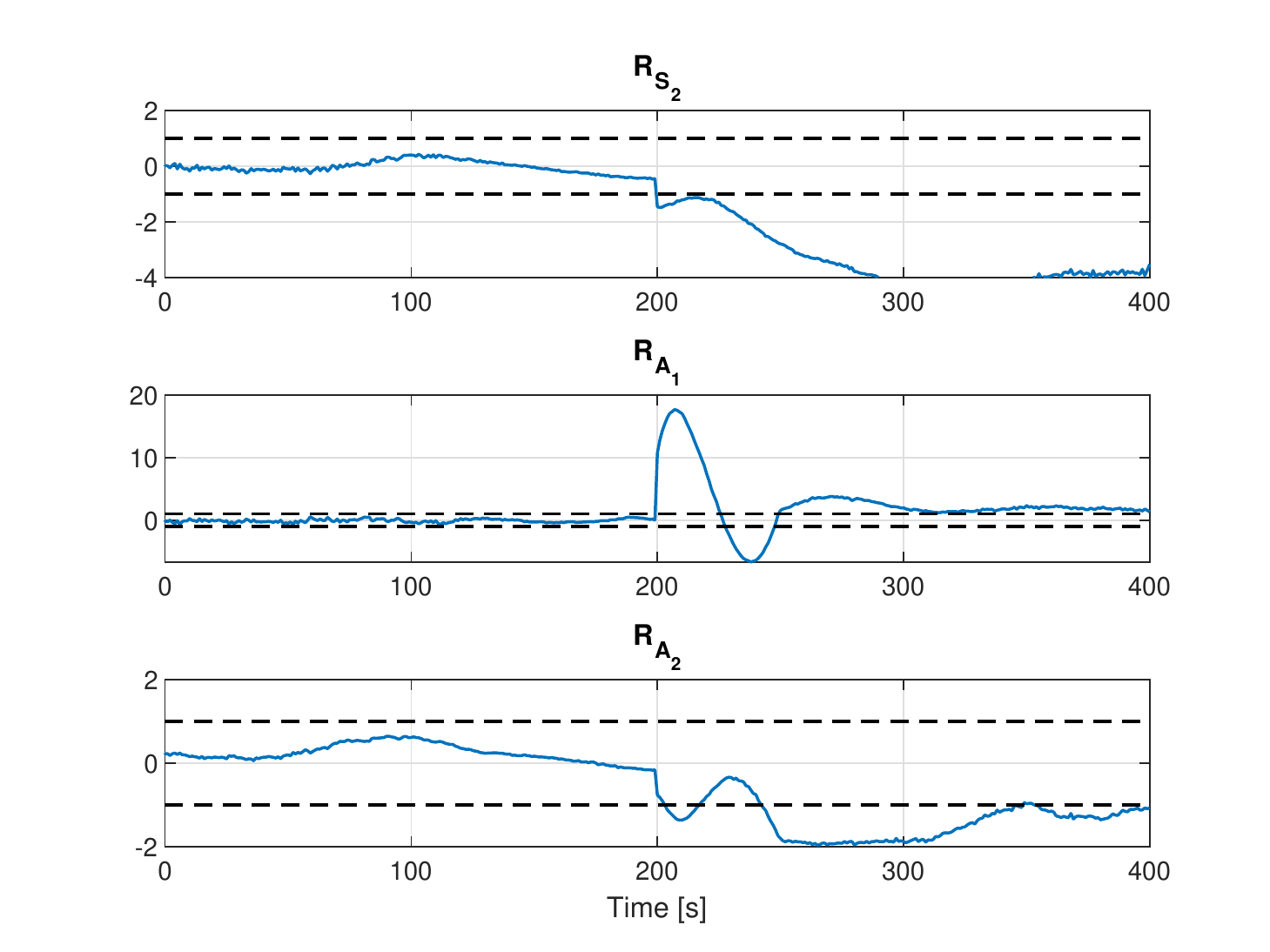}
\caption{\textit{Case A: residues responses to a fault on sensor $\mathsf{S}_1$.}}
\label{A_exp_fault_T1}
\end{figure}

\begin{figure}[htb!]
\centering
\includegraphics[width=\columnwidth]{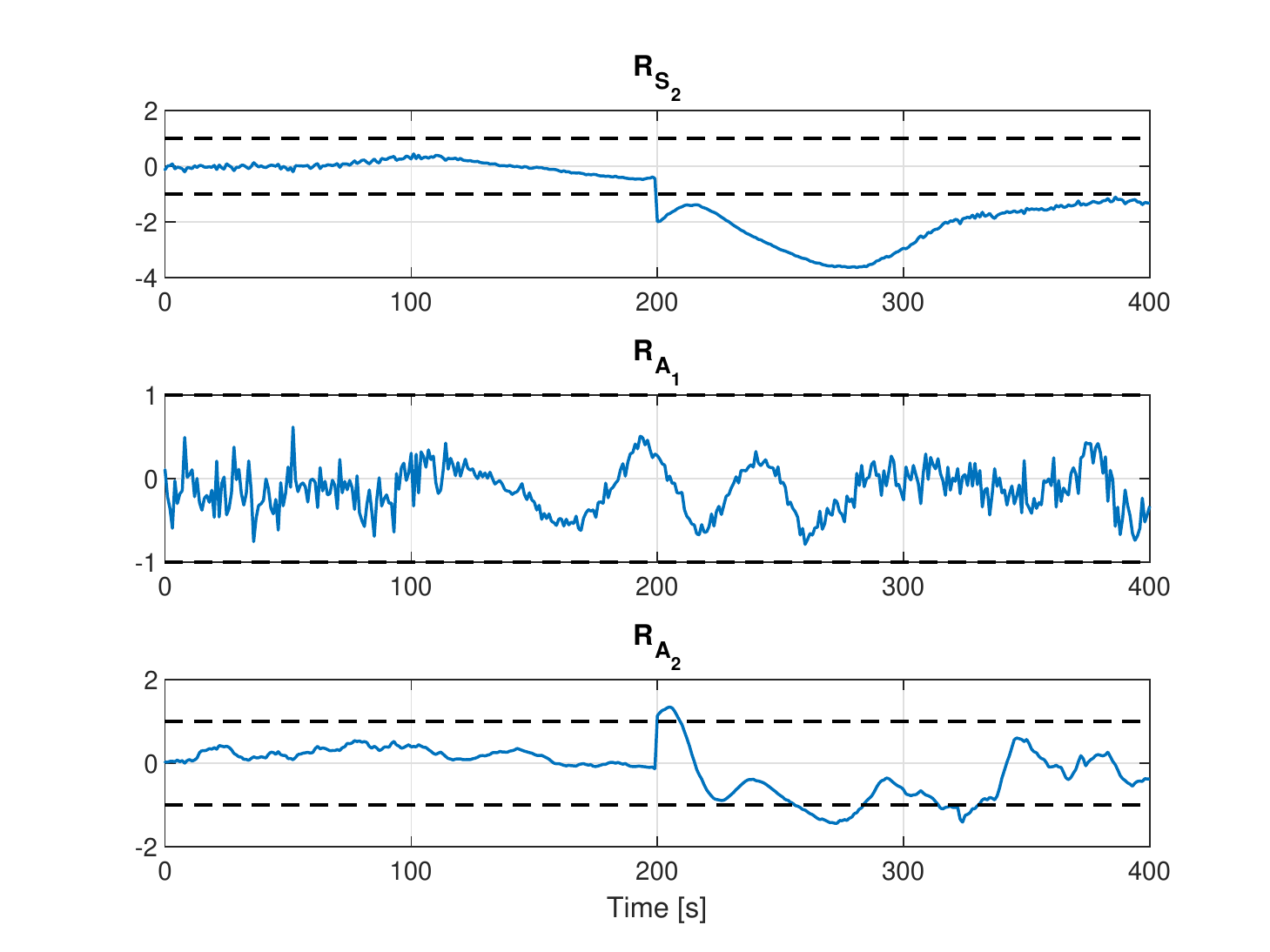}
\caption{\textit{Case A: residues responses to a fault on the sensor $\mathsf{S}_2$.}}
\label{A_exp_fault_T2}
\end{figure}

\begin{figure}[htb!]
\centering
\includegraphics[width=\columnwidth]{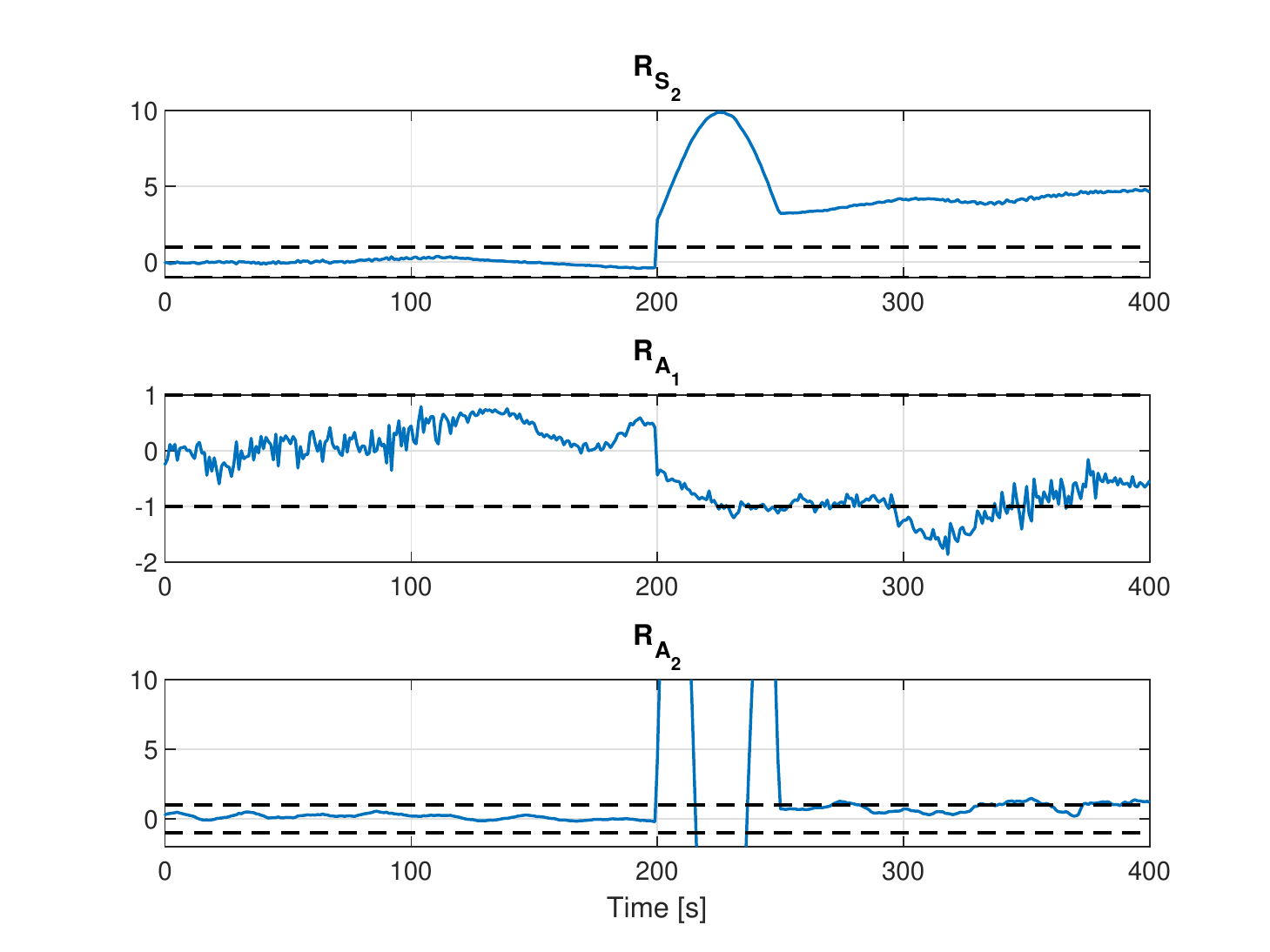}
\caption{\textit{Case A: residues responses to a fault on the sensor $\mathsf{S}_3$.}}
\label{A_exp_fault_T3}
\end{figure}

\begin{figure}[htb!]
\centering
\includegraphics[width=\columnwidth]{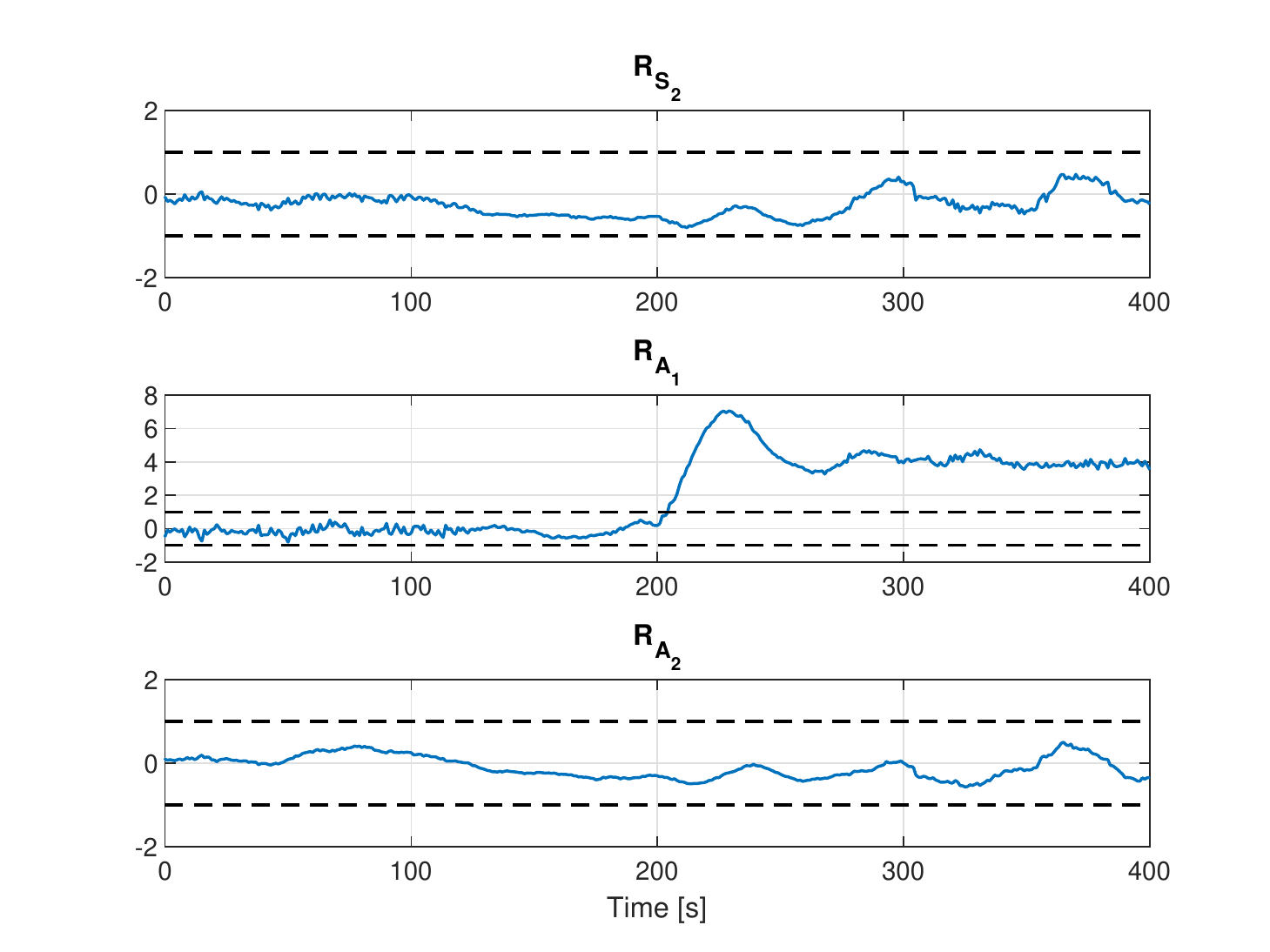}
\caption{\textit{Case A: residues responses to a fault on actuator $\mathsf{A}_1$.}}
\label{A_exp_fault_A1}
\end{figure}

\begin{figure}[htb!]
\centering
\includegraphics[width=\columnwidth]{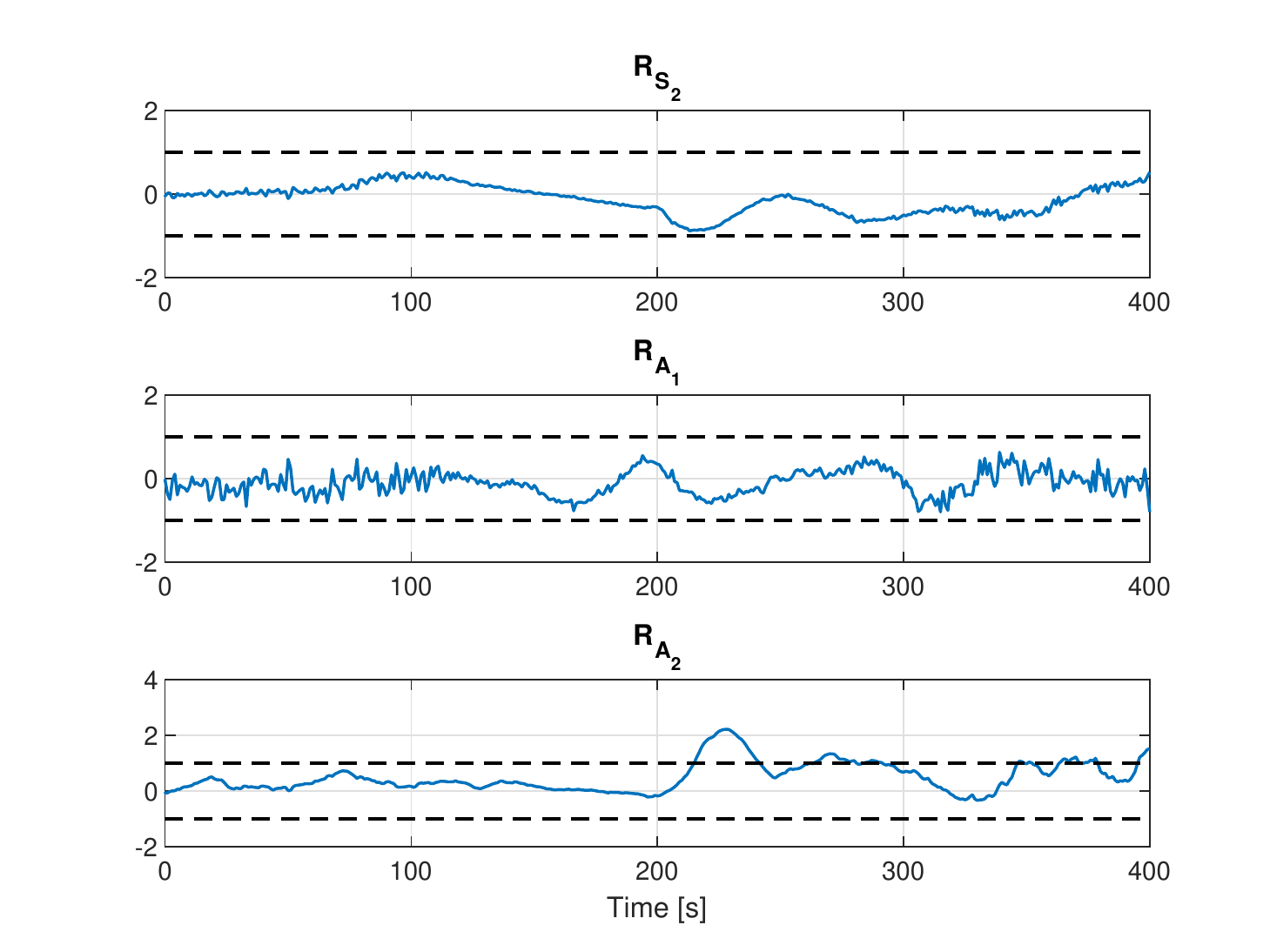}
\caption{\textit{Case A: residues responses to a fault on actuator $\mathsf{A}_2$.}}
\label{A_exp_fault_A2}
\end{figure}

For multiplicative faults we consider a $20\%$ failure for sensors and actuators. At time $t=200\, \SI{}{s}$, sensors measure $80\%$ of the actual water level measurements instead of $100\%$, and for actuators, a $20\%$ failure is considered:
\begin{itemize}
\item[--] Figures~\ref{A_exp_fault_T1} and \ref{A_exp_fault_T2} show the residues with their respective thresholds for one single fault on sensors $\mathsf{S}_1$ and $\mathsf{S}_2$, respectively. As it can be observed, the first two fault signature columns $\Sigma_1$ and $\Sigma_2$ in \eqref{sig_z1_2} are validated.

\item[--] All residues are affected by a fault on the sensor $\mathsf{S}_3$, however, as mentioned earlier (see also Table~\ref{table_sens_Z1}), residue $R_{A_1}^{Z_1}$ is weakly sensitive to this fault, as can be seen in Figure~\ref{A_exp_fault_T3}, which validates the definition of the fault signature column $\Sigma_3$ in \eqref{sig_z1_2}. 

\item[--] Figures~\ref{A_exp_fault_A1} and Figure~\ref{A_exp_fault_A2} show the residues and their thresholds for faults on actuators $\mathsf{A}_1$ and $\mathsf{A}_2$, respectively. These results validate the definition of the fault signature columns $\Sigma_4$ and $\Sigma_5$ in \eqref{sig_z1_2}.
\end{itemize}%

According to Definition~\ref{detectability}, all faults on the system sensors and actuators are detectable. In addition, since fault alarm signatures $\Sigma_1$, $\Sigma_4$ and $\Sigma_5$ are distinct, faults on sensor $\mathsf{S}_1$ and actuators $\mathsf{A}_1$ and $\mathsf{A}_2$ are isolable, according to Definition~\ref{isolability}. This reflects the fact that if, at some point during system operation, a fault alarm is launched with the signature $\Sigma_1$ then we conclude that the sensor $\mathsf{S}_1$ is faulty. Nevertheless, if we obtain a signature like $\Sigma_2$, the fault could be on the sensor $\mathsf{S}_2$ or $\mathsf{S}_3$, since signatures $\Sigma_2$ and $\Sigma_3$ are identical, hence, a fault on $\mathsf{S}_2$ or $\mathsf{S}_3$ cannot be isolated. Therefore, the number of distinct fault alarm signatures is $\mu=3<~p+m$.

\subsubsection*{Case B: using two flat outputs} 
In order to improve the isolability, a second flat output is needed. For the sequel, a second flat output $z=(x_2,x_3)^T$ is used. In the following, we denote by $Z_1$ the first flat output $Z_1=(z_{11}^s,z_{12}^s)^T=(x_1^s,x_3^s)^T$ and by $Z_2$ the second flat output $Z_2=(z_{21}^s,z_{22}^s)^T=(x_2^s,x_3^s)^T$. The signature matrix associated to $Z_1$ is given by \eqref{sig_z1_2} and we denote it by $\mathbf{S}_1$. The  number of distinct signatures is $\mu_1=3<p+m$.

In order to construct the signature matrix $\mathbf{S}_2$, associated to $Z_2$, the redundant inputs and outputs are first computed using \eqref{xz,uz} and \eqref{virtual-output}:
\begin{align}
y_1^{Z_2}&=z_{22}^s+\frac{1}{\mu_{13}^2}\Big(\dot{z}_{22}^s+\mu_{32}\sqrt{z_{22}^s-z_{21}^s}\Big)^2 \nonumber\\
y_2^{Z_2}&=z_{21}^s \nonumber\\
y_3^{Z_2} & =z_{22}^s \\
u_1^{Z_2}& =\dot{y}_{1}^{Z_2}+\mu_{13}\sqrt{y_1^{Z_2}-z_{22}^s}\nonumber\\
u_2^{Z_2} &=\dot{z}_{21}^s+\mu_{20}\sqrt{z_{21}^s}-\mu_{32}\sqrt{z_{22}^s-z_{21}^s}\nonumber
\end{align}
According to Remark~\ref{rem1}, residues associated to sensors $\mathsf{S}_2$ and $\mathsf{S}_3$ are identically zero. Then, the vector of residues is truncated to:
\begin{align}
r_\tau^{Z_2}=\begin{pmatrix}
R_{\mathsf{S}_1}^{Z_2}\\  R_{\mathsf{A}_1}^{Z_2} \\ R_{\mathsf{A}_2}^{Z_2} 
\end{pmatrix}=\begin{pmatrix}
 y_{2}^s \\  u_{1} \\ u_{2} 
\end{pmatrix}- \begin{pmatrix}
 y_2^{Z_2} \\  u_1^{Z_2} \\ u_2^{Z_2}
\end{pmatrix}.
\label{31}
\end{align}

\begin{table*}[htb!]
    \centering
    \begin{tabular}{|c|c|c|c|c|c|c|c|c|c|}
    \hline
         & $\pdv{y_1^s}$ & $\pdv{y_2^s}$ & $\pdv{y_3^s}$ & $\pdv{\dot{y}_1^s}$ & $\pdv{\dot{y}_2^s}$& $\pdv{\dot{y}_3^s}$ & $\pdv{\ddot{y}_3^s}$ & $\pdv{u_1}$ & $\pdv{u_2}$ \\ \hline
        $R_{S_1}^{Z_2}$ & 1 & 1 & 2 & 0 & 0 & 5262 & 0 & 0 & 0 \\ \hline
        $R_{A_1}^{Z_2}$ & 1 & $10^{-4}$ & $10^{-4}$ & 0 & 1 & 3 & 5262 & 1 & 0  \\ \hline
        $R_{A_2}^{Z_2}$ & 0 & $10^{-4}$ & $10^{-4}$ & 0 & 1 & 0 & 0 & 0 & 1  \\ \hline
    \end{tabular}
    \caption{Partial derivatives of residues $R_{S_2}^{Z_2}$, $R_{A_1}^{Z_2}$, and $R_{A_2}^{Z_2}$ with respect to $\zeta$ and its derivatives computed at the equilibrium point $x_{1_e}=0.20$~m, $x_{2_e}=0.10$~m, $x_{3_e}=0.15$~m.}
    \label{table_sens_Z2}
\end{table*}

As for the flat output $Z_1$ (see Table~\ref{table_sens_Z1}), the numerical values for the partial derivatives at the equilibrium state $x_{1_e}=0.20$~m, $x_{2_e}=0.10$~m, $x_{3_e}=0.15$~m are given in Table~\ref{table_sens_Z2}.

Using Definition~\ref{def_signature} and the threshold $\mathbf{Th}=0.5$, the signature matrix associated to $Z_2$ is obtained as:
\begin{equation}
\mathbf{S}_2=\begin{pmatrix}
1 & 1 & 1 & 0 & 0\\
1 & 1 & 1 & 1 & 0\\
0 & 1 & 0 & 0 & 1
\end{pmatrix}
\label{32}
\end{equation}
and faults on sensors $\mathsf{S}_1$ and $\mathsf{S}_3$ are not isolable.
The number of distinct signature is $\mu_2=3<p+m$.

In order to prove the independence of the flat outputs $Z_1$ and $Z_2$, the following augmented signature matrix is constructed using the experimental results:
\begin{equation}
\widetilde{\mathbf{S}}=\begin{pmatrix}
\mathbf{S}_1\\ \mathbf{S}_2
\end{pmatrix}=\begin{pmatrix}
1 & 1 & 1 & 0& 0\\
1 & 0 & 0 & 1 & 0\\
1 & 1 & 1 & 0 & 1 \\
1 & 1 & 1 & 0 & 0\\
1 & 1 & 1 & 1 & 0\\
0 & 1 & 0 & 0 & 1
\end{pmatrix}.
\label{Stilde1}
\end{equation}
The number of distinct signature of $\widetilde{\mathbf{S}}$ is $\widetilde{\mu}=5$, thus, even in the presence of uncertainties, the condition \eqref{condition} is satisfied, then the flat outputs $Z_1$ and $Z_2$ are independent. In addition, $\widetilde{\mu}=p+m$, then, according to Definition~\ref{def_independence}, the flat outputs $Z_1$ and $Z_2$ provide full isolability of faults. 

This result is demonstrated experimentally:
\begin{itemize}
\item[--] Figure~\ref{B_exp_fault_T1} shows that if a fault affects the sensor $\mathsf{S}_1$, all the residues exceed their threshold except $R_{\mathbf{A}_2}^{Z_2}$ which confirms the signature $\Sigma_1$ of the matrix $\widetilde{\mathbf{S}}$.
\begin{figure}[htb!]
\centering
\includegraphics[width=\columnwidth]{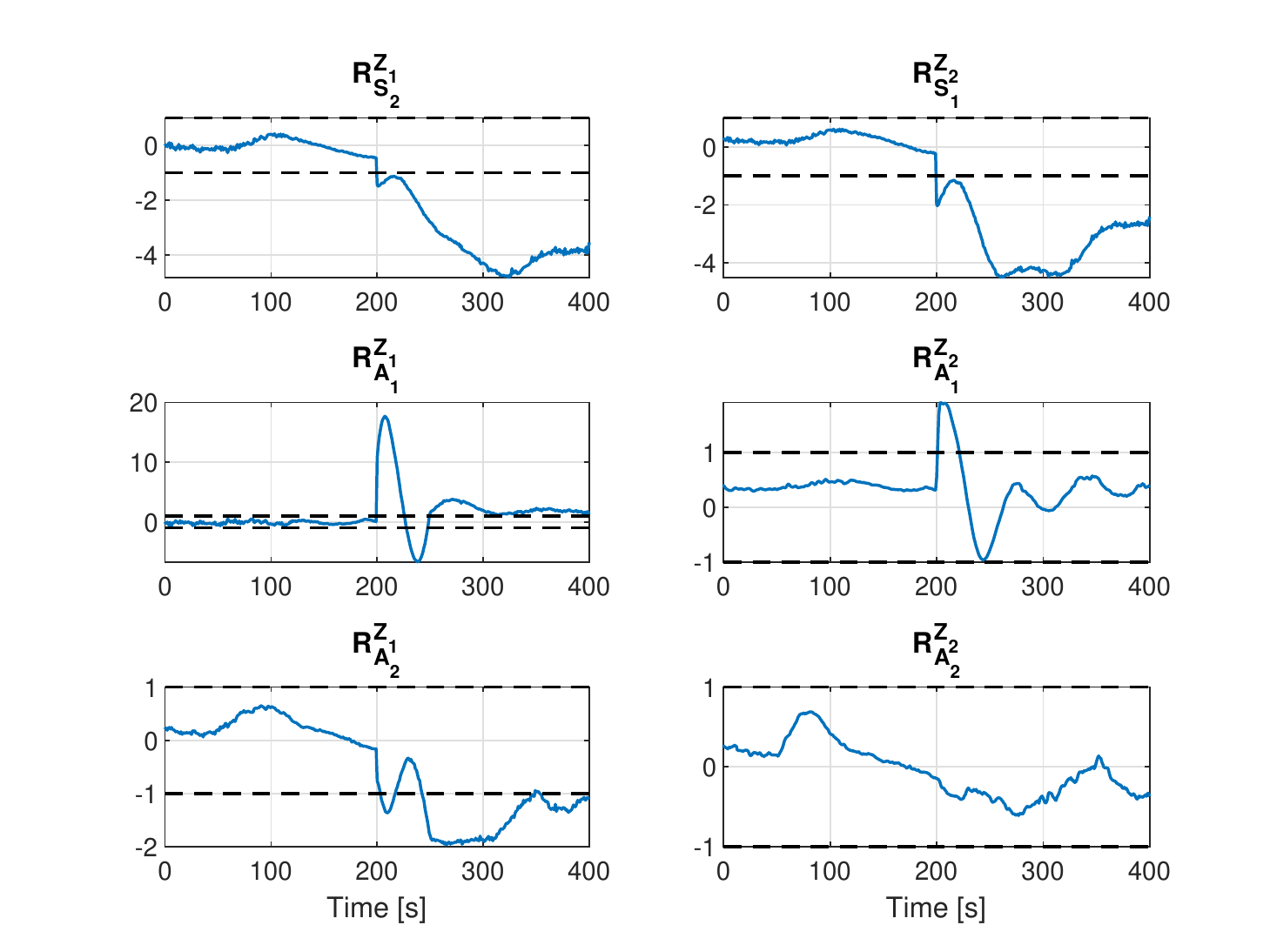}
\caption{\textit{Case B: residues responses to a fault on sensor $\mathsf{S}_1$.}}
\label{B_exp_fault_T1}
\end{figure}

\item[--] Figure~\ref{B_exp_fault_T2} shows that if a fault affects sensor $\mathsf{S}_2$, all the residues exceed their threshold except $R_{\mathbf{A}_1}^{Z_1}$ which confirms the signature $\Sigma_2$ of the matrix $\widetilde{\mathbf{S}}$.

\begin{figure}[htb!]
\centering
\includegraphics[width=\columnwidth]{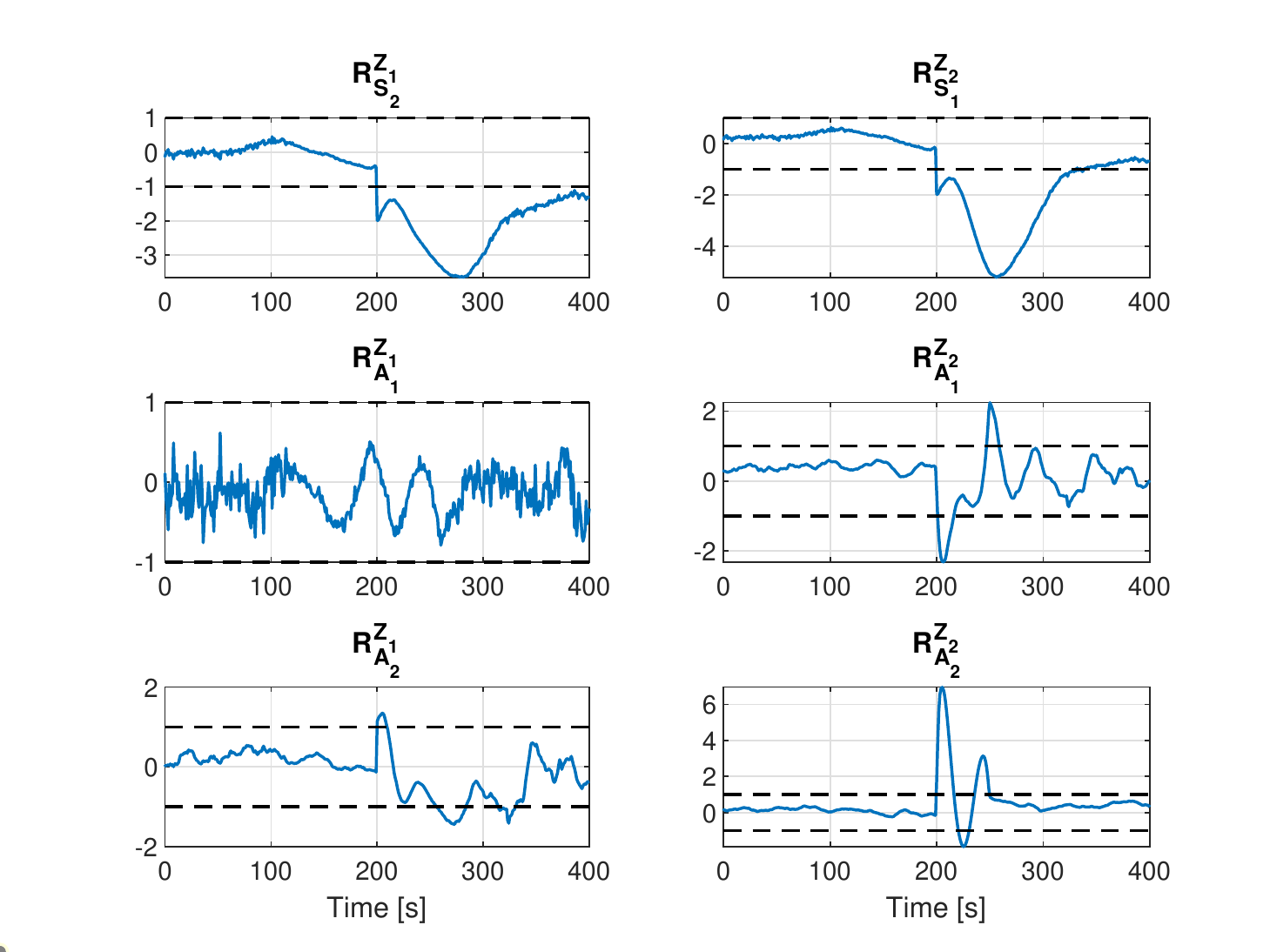}
\caption{\textit{Case B: residues responses to a fault on sensor $\mathsf{S}_2$.}}
\label{B_exp_fault_T2}
\end{figure}

\item[--] Finally, Figure~\ref{B_exp_fault_T3} confirms that residues $R_{\mathbf{A}_1}^{Z_1}$ and $R_{\mathbf{A}_2}^{Z_2}$ are weakly affected by a fault on the sensor $\mathsf{S}_3$.
\begin{figure}[htb!]
\centering
\includegraphics[width=\columnwidth]{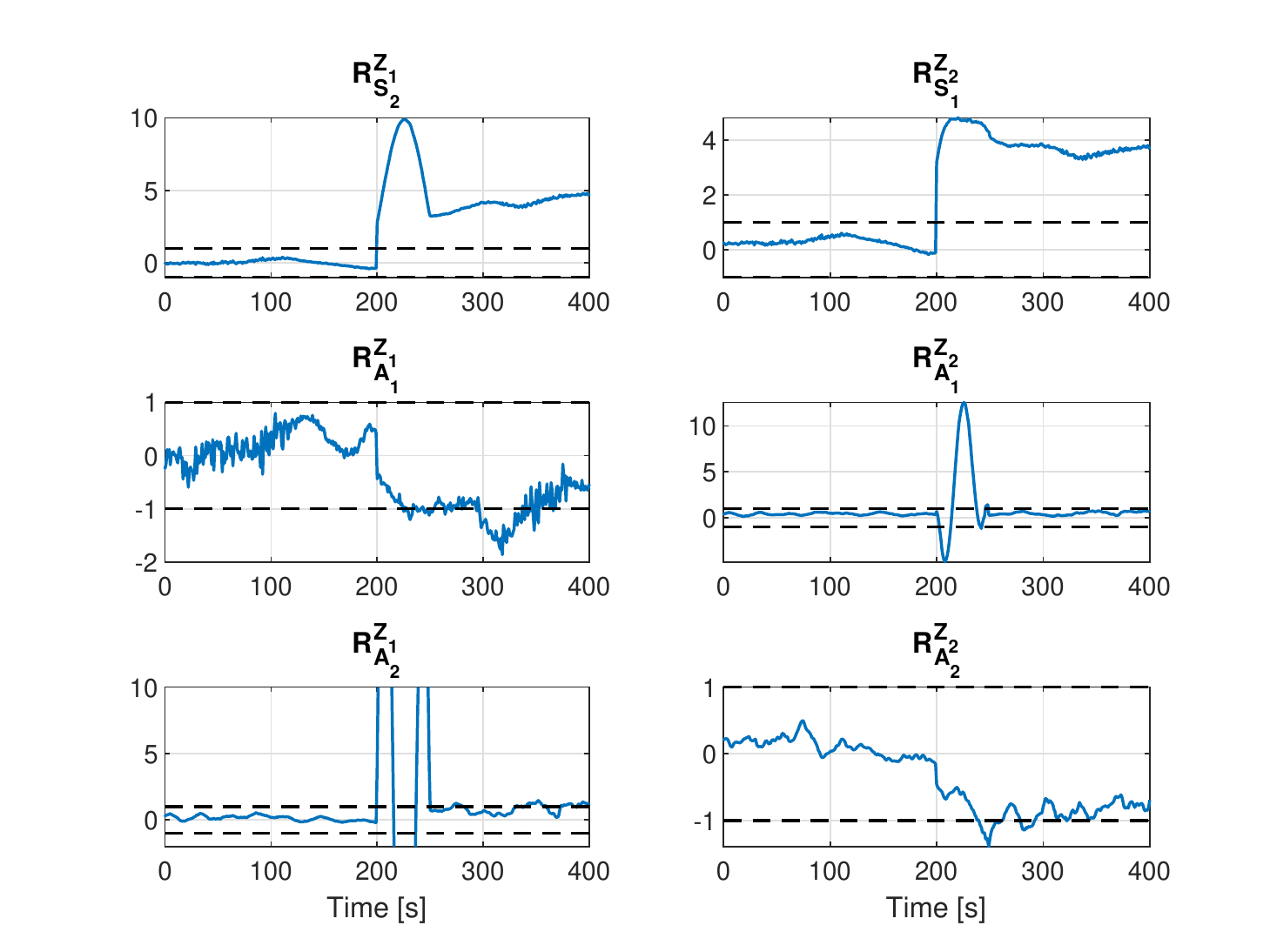}
\caption{\textit{Case B: residues responses to a fault on sensor $\mathsf{S}_3$.}}
\label{B_exp_fault_T3}
\end{figure}
\end{itemize}

\section{Conclusion}
\label{section6}
In this paper, a FDI method based on the flatness property of nonlinear systems is presented. The flat output measurement is used to calculate the redundant variables and then generate the residues. Moreover, it has been shown that sometimes, using a single flat output is not sufficient to ensure full isolability and several flat outputs may be needed. These flat outputs must be independent in the sense that by using them together the number of isolable faults increases. Therefore, a full presentation of this flat output characterization is provided.  sensitivity analysis of residues with respect to faults is proposed in order to robustly design  the fault signature matrix. Finally, the validity of this method has been shown experimentally on the DTS200 three-tank benchmark.

\section*{Acknowledgments:}
The authors gratefully acknowledge Jean Lévine, professor emeritus in CAS, Ecole des Mines de Paris, for his assistance in the theoretical part of this paper, as well as the reviewers for their valuable suggestions. 
\bibliography{library}

\end{document}